\documentclass[pra,twocolumn,preprintnumbers,amsmath,amssymb,superscriptaddress]{revtex4}

\usepackage[T1]{fontenc}
\usepackage[latin1]{inputenc}
\usepackage[english]{babel}
\usepackage{graphicx,subfigure,color}
\usepackage{dcolumn}
\usepackage{bm,bbold,braket}
\usepackage{relsize}
\usepackage{subfigure}
\usepackage{amsmath,amssymb,amstext}
\usepackage[]{todonotes}

\newcommand{\vect}[1]{\ensuremath{\bm{#1}}}
\newcommand{\eq}[1]{\begin{equation} #1 \end{equation}}
\newcommand{\eqa}[1]{\begin{eqnarray} #1 \end{eqnarray}}

\newcommand{\ud}{\mathrm{d}}
\newcommand{\ue}{\mathrm{e}}

\newcommand{\Ho}{\hat{H}}
\newcommand{\no}{\hat{n}}
\renewcommand{\ao}{\hat{a}^{\phantom{\dag}}}
\renewcommand{\aa}{\hat{a}^\dag}
\newcommand{\rd}{{\mathrm d}}
\newcommand{\ri}{{\mathrm i}}
\newcommand{\re}{{\mathrm e}}
\newcommand{\la}{\langle}
\newcommand{\ra}{\rangle}
\newcommand{\Uo}{\hat{U}}

\newcommand{\bq}{{\bm q}}
\newcommand{\br}{{\bm r}}

\newcommand{\bF}{{\bm F}}

\newcommand{\bk}{{\bm k}}

\newcommand{\dw}{\downarrow}
\newcommand{\up}{\uparrow}
\newcommand{\atan}{\mathrm{atan}}

\newcommand{\etal}{\textit{et al.,} }

\begin{document}
\title{Non-Abelian gauge fields and topological insulators in shaken optical lattices}

\author{Philipp Hauke}
    \email{philipp.hauke@icfo.es}
    \affiliation{ICFO -- Institut de Ci\`{e}ncies Fot\`{o}niques, Parc Mediterrani de la Tecnologia, E-08860 Castelldefels, Spain}
\author{Olivier Tieleman}
    \affiliation{ICFO -- Institut de Ci\`{e}ncies Fot\`{o}niques, Parc Mediterrani de la Tecnologia, E-08860 Castelldefels, Spain}
\author{Alessio Celi}
    \affiliation{ICFO -- Institut de Ci\`{e}ncies Fot\`{o}niques, Parc Mediterrani de la Tecnologia, E-08860 Castelldefels, Spain}
\author{Christoph \"Olschl\"ager}
    \affiliation{Institut f\"ur Laserphysik, Universit\"at Hamburg, Luruper Chaussee 149, D-22761 Hamburg, Germany}
\author{Juliette Simonet}
    \affiliation{Institut f\"ur Laserphysik, Universit\"at Hamburg, Luruper Chaussee 149, D-22761 Hamburg, Germany}
\author{Julian Struck}
    \affiliation{Institut f\"ur Laserphysik, Universit\"at Hamburg, Luruper Chaussee 149, D-22761 Hamburg, Germany}
\author{Malte Weinberg}
    \affiliation{Institut f\"ur Laserphysik, Universit\"at Hamburg, Luruper Chaussee 149, D-22761 Hamburg, Germany}
\author{Patrick Windpassinger}
    \affiliation{Institut f\"ur Laserphysik, Universit\"at Hamburg, Luruper Chaussee 149, D-22761 Hamburg, Germany}
\author{Klaus Sengstock}
    \affiliation{Institut f\"ur Laserphysik, Universit\"at Hamburg, Luruper Chaussee 149, D-22761 Hamburg, Germany}
\author{Maciej Lewenstein}
    \affiliation{ICFO -- Institut de Ci\`{e}ncies Fot\`{o}niques, Parc Mediterrani de la Tecnologia, E-08860 Castelldefels, Spain}
    \affiliation{ICREA -- Instituci{\'o} Catalana de Recerca i Estudis Avan\c{c}ats, Lluis Companys 23, E-08010 Barcelona, Spain}
\author{Andr\'e Eckardt}
    \affiliation{Max-Planck-Institut f\"ur Physik komplexer Systeme, N\"othnitzer Str.\ 38, D-01187 Dresden, Germany}

\date{\today}

\begin{abstract}
Time-periodic driving like lattice shaking offers a low-demanding method to 
generate artificial gauge fields in optical lattices. We identify the relevant 
symmetries that have to be broken by the driving function for that purpose and 
demonstrate the power of this method by making concrete proposals for its 
application to two-dimensional lattice systems: We show how to tune 
frustration and how to create and control band touching points like Dirac
cones in the shaken kagom\'e lattice. We propose the realization of a 
topological and a quantum spin Hall insulator in a shaken spin-dependent
hexagonal lattice. We describe how strong artificial magnetic fields can be
achieved for example in a square lattice by employing superlattice modulation.
Finally, exemplified on a shaken spin-dependent square lattice, we develop a 
method to create strong non-Abelian gauge fields. 
\end{abstract}

\pacs{PACS}

\maketitle

Topological order and topological insulators \cite{Hasan2010} are
currently in the center of interest of quantum physics, especially because of 
their possible applications in quantum information and spintronics 
\cite{Nayak2008}. 
For this reason, there is an ongoing search for feasible realizations of such systems in- and outside of solid-state physics.
Here, ultracold 
ground-state atoms provide a promising playground \cite{Lewenstein2012} 
(although Rydberg-excited atoms \cite{Weimer2010}, trapped ions 
\cite{Barreiro2011}, and photons in nano-structured materials 
\cite{Kitagawa2011etal} offer interesting alternatives).
Typically, topological effects require ultra-strong gauge fields or 
spin-orbit-like couplings. There are several ways to achieve these with 
ultracold atoms, from trap rotation \cite{Fetter2009},
microrotation \cite{Sorensen2005etal}, to Berry phase imprinting 
\cite{Dalibard2011etal}. In optical lattices, combining laser-induced 
tunneling with superlattice techniques allows for strong Abelian
\cite{Jaksch2003} and non-Abelian \cite{Osterloh2005} gauge-fields 
and for the realization of topological insulators \cite{Mazza2012}. 
So far, the first lattice experiments led to the creation of staggered flux 
lattices \cite{Aidelsburger2011}. Many other groups follow
this direction of research \cite{Tarruell2011etal}. 

Recently, there has been a burst of interest in another, experimentally less 
demanding, approach, namely periodic lattice shaking.  
Sinusoidal shaking leads to a change of strength, or even sign of the 
tunneling and allows to control the Mott-insulator--superfluid transition 
\cite{Eckardt2005,Pisa} (for a recent work in hexagonal geometry, see 
\cite{Koghee2012}).
While in the square lattice this introduces neither frustration nor synthetic 
gauge fields, in the triangular lattice a sign-change of the tunneling is 
equivalent to a $\pi$-flux Abelian field \cite{KalmeyerLaughlin1987}. 
Such a system mimics frustrated antiferromagnetism, classical for weakly 
interacting bosons \cite{Struck2011}, and quantum in the hard-core boson limit 
\cite{Eckardt2010}, where it is expected to exhibit exotic spin-liquid phases 
\cite{Schmied2008etal}. 
Recently, it was demonstrated that by breaking temporal symmetries of
the shaking trajectory, one can create phases of the tunneling
in an optical lattice \cite{Struck2012,Sacha2012} (see also 
Ref.~\cite{JimenezGarcia2012}), and that in this way tunable Abelian fluxes 
through triangular plaquettes may be generated \cite{Struck2012}. 
In this letter, we discuss non-trivial generalizations of this approach 
that involve also AC-induced tunneling and spinful
particles. This allows us to simulate Abelian and even non-Abelian
SU(2) gauge-fields in different lattice geometries, as well as topological
insulators. 
To this, we employ non-standard optical lattices, like kagom\'e and spin-dependent square and
hexagonal lattices, and consider scenarios based on superlattice modulation.

\paragraph*{Basic scheme, and temporal symmetries.} 
We consider a system of ultracold atoms in a driven optical lattice described 
by the Hubbard Hamiltonian $\Ho(t)=-\sum_{\la ij\ra}J_{ij}\aa_i\ao_j 
+ \sum_i v_i(t)\no_i +\Ho_\text{os}$ with (bare) tunneling matrix elements
$J_{ij}$ and annihilation and number operators $\ao_i$ and $\no_i$ for
particles (bosons or fermions) at site $i$; $\Ho_\text{os}$ collects 
on-site terms describing interactions or a weak static potential.
The potential $v_i(t)=v^\omega_i(t)+\nu_i\hbar\omega$ consists of two parts: a
rapid periodic drive $v^\omega_i(t)=v^\omega_i(t+T)$ of frequency 
$\omega=2\pi/T$ and zero time average $\la v^\omega_i(t)\ra_T =0$ with $\la\cdot\ra_T\equiv \frac{1}{T}\int_0^T\!\cdot\rd t$; and (unlike in
Ref.~\cite{Struck2012}) strong static energy offsets $\nu_i\hbar\omega$ with 
integers $\nu_i$. For $\hbar\omega\gg J_{ij}$ a large energy difference 
$\nu_{ij}\hbar\omega\ne0$ (here and below we use the double-index shorthand 
$x_i-x_j\equiv x_{ij}$) practically prohibits tunneling between $i$ and $j$, 
unless the resonant periodic driving leads to AC-induced tunneling (ACT) 
\cite{Eckardt2007}, as it has been observed in recent experiments 
\cite{Sias2008etal}. Later on, we will augment our model by a spin degree of
freedom $s=\up,\dw$. 

A gauge transformation $\Uo=\exp\left(\ri\sum_i\chi_i(t)\no_i\right)$, where
$\chi_i(t) = \chi^\omega_i(t)-\nu_i\omega t +\gamma_i$ with
$\hbar\chi^\omega_i(t)=-\int_0^t\ud\tau v^\omega_i(\tau)+\la\int_0^t \ud\tau 
v^\omega_i(\tau)\ra_T$ and constants $\gamma_i$, leads to the new Hamiltonian 
$\Ho'(t)=\Uo^\dag\Ho\Uo-\ri\hbar\Uo^\dag(\rd_t\Uo)$, 
which can be approximated by its time average $\Ho_\text{eff}\equiv-\sum_{\la 
ij\ra}J_{ij}^\text{eff}\aa_i\ao_j +\Ho_\text{os}$ if $\hbar\omega$ is large 
compared to both the $J_{ij}$ and the energy scales 
of $\Ho_\text{os}$. In this treatment, the initial energy offsets 
$\nu_{i}\hbar\omega$ enter via the effective tunneling matrix elements 
$J^\text{eff}_{ij}=J_{ij} \la \re^{-\ri(\chi_{ij}(t))} \ra_T$ only, 
and in $\Ho_\text{eff}$ all sites appear to have the same energy.
In the undriven system, for $\nu_{ij}\ne0$, the large energy difference $\nu_{ij}\hbar\omega$ suppresses tunneling between sites $i$ and $j$, 
and this fact is reflected in $\Ho_\text{eff}$ by a vanishing effective tunneling $J^\text{eff}_{ij}=0$ at vanishing driving $v^\omega_{ij}=0$. 
In turn, finite driving $v^\omega_{ij}\neq0$ can establish 
coherent ACT with $J_{ij}^\text{eff}\ne0$, where the energy difference 
$\nu_{ij}\hbar\omega$ is bridged by $\nu_{ij}$ quanta $\hbar\omega$.

The leitmotif of the present work 
is to use this control scheme to induce Peierls-type phases 
\begin{equation}\label{eq:peierls}
\theta_{ij}
=\arg\left(\la\re^{-\ri(\chi^\omega_{ij}(t) -\nu_{ij}\omega t +\gamma_{ij}
)}\ra_T\right)
\end{equation}
that cannot be eliminated globally by choice of gauge, i.e., by adjusting the 
constants $\gamma_i$. 
Such non-trivial phases correspond to artificial Abelian gauge fields; 
the gauge-invariant magnetic flux 
$\phi_P\in(-\pi,\pi]$ piercing a lattice plaquette $P$ is (modulo $2\pi$) 
obtained by summing the $\theta_{ij}$ around $P$. 
We find that the
\emph{global} reflection symmetry (r) $v^\omega_i(-t-\tau)=v^\omega_i(t-\tau)$
with respect to a global time $\tau$ (using the choice 
$\gamma_i=-\nu_i\omega\tau$)
implies trivial $\theta_{ij}=0$.
Moreover, 
if ACT is not involved ($\nu_{ij}=0$), $\theta_{ij}=0$ follows already 
from the \emph{local} reflection symmetry (r') 
$v^\omega_{ij}(-t-\tau_{ij})=v^\omega_{ij} (t-\tau_{ij})$ 
with independent local times $\tau_{ij}$ (since 
$\gamma_{ij}=\nu_{ij}\omega\tau=0$, independent of $\tau$), or from the shift 
antisymmetry (s)
$v^\omega_i(t-\frac{T}{2})=-v^\omega_i(t)$ (choosing $\gamma_i=0$) 
\cite{FootnoteRatchets}. Therefore, ACT significantly reduces the constraints 
on the driving function $v^\omega_i(t)$ for the creation of artificial 
magnetic fields. This is nicely exemplified by recent proposals 
where already simple sinusoidal forcing [fulfilling (r') and (s)] leads to 
magnetic fields when combined with ACT  -- provided the temporal phase of the 
driving can be made site dependent [thus breaking (r)] \cite{Kolovsky2011}. 
In the following, we consider experimentally feasible scenarios where the 
whole system is driven in phase [such that both (r) and (s) are broken]. 

We will later generalize the scheme described in the two preceding paragraphs to the case of spin-1/2 particles and show how non-Abelian gauge fields can be realized.

\begin{figure}[t]\centering
\includegraphics[width = 1\linewidth]{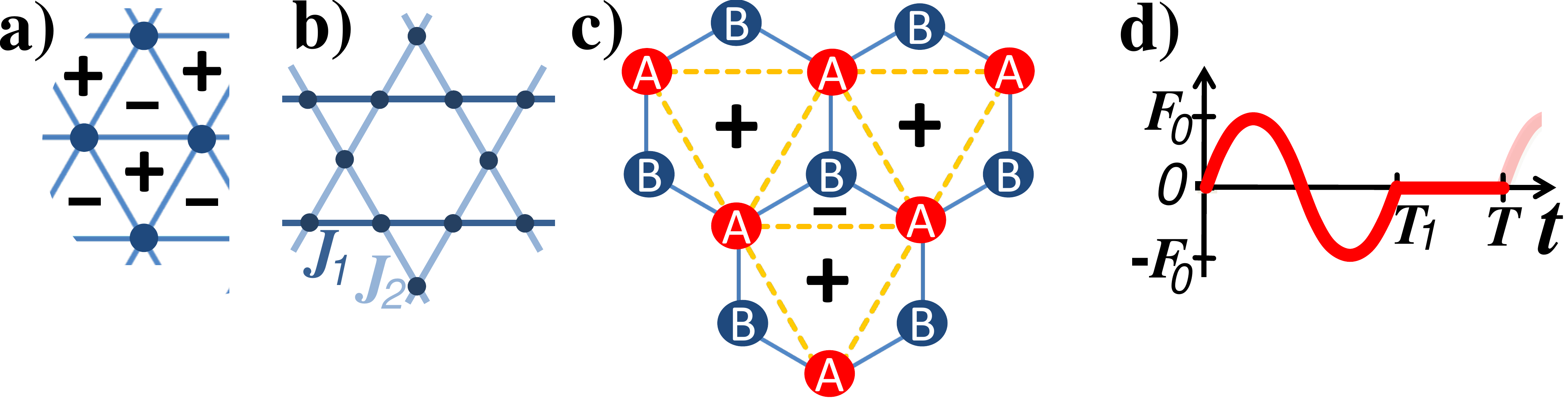}
\caption{\label{fig:tri} (color online) {(a-c)} Lattice geometries
involving triangular plaquettes pierced by an artificial magnetic 
flux $\phi_{\nabla,\Delta}=\pm\phi$ (indicated by $+$ and $-$): 
{(a)} triangular lattice, {(b)} kagom\'e lattice with tunneling $J_1$ ($J_2$) along the darker horizontal (lighter diagonal) bonds, and {(c)} hexagonal lattice 
with nearest-neighbor ACT (solid lines) between shallow 
A- and deep B-sites and next-nearest-neighbor tunneling between 
A-sites (dashed lines). (d) Driving function breaking symmetries (r) and (s).}
\end{figure}

\paragraph*{Homogeneous forcing and triangular plaquettes.} 
Let us consider a homogeneous time-periodic force $\bF(t)$, such as an
inertial force created by shaking the lattice along a periodic orbit.
For $\nu_i=0$, the driving potential $v^\omega_i(t)=-\br_i\cdot\bF(t)$ (with 
site position $\br_i$) results in Peierls phases $\theta_{ij}$ that only depend 
on the vector $\br_{ij}=\br_{i}-\br_{j}$ connecting the two sites $i$ and $j$, $\theta_{ij}=f(\br_{ij})$. Using
Eq.~(\ref{eq:peierls}), one finds that $f(-\br_{ij})=-f(\br_{ij})$ and, 
therefore, homogeneous forcing cannot be used to create artificial magnetic 
fluxes through plaquettes with pairwise parallel edges. Since, however, generically 
$\theta_{ij}$ depends in a non-linear fashion on $\br_{ij}$ [$f(\br_{ij})$ is 
not of the form ${\bm b}\cdot\br_{ij}$], one can use lattice shaking to induce a strong and tunable artificial 
magnetic flux $\phi_\nabla$ through, e.g., a downwards pointing triangular plaquette $\nabla$.
In the supplemental material \cite{supp}, we analytically compute this flux for unidirectional 
forcing.
The inversion of the triangular plaquette $\nabla\to\Delta$ reverses the sign 
of the flux, $\phi_\Delta=-\phi_\nabla$, such that staggered fluxes can be 
achieved in the triangular or kagom\'e lattice as shown in Fig.~\ref{fig:tri}a 
and b. 
Since these flux configurations stem from homogeneous forcing they do not 
break the translational symmetry of the lattice. 

Tuning the staggered flux allows one to continuously control the degree of 
frustration in these lattices from none for zero-flux to maximum for 
$\pi$-flux [corresponding to ferromagnetic ($-J^\text{eff}_{ij}<0$) and 
antiferromagnetic coupling ($-J^\text{eff}_{ij}>0$), respectively].  
The fully-frustrated regime gives rise to intriguing physics. For example, the 
flat lowest band of the kagom\'e lattice makes the system extremely 
susceptible towards interaction-driven physics \cite{Huber2010}; moreover, the 
case of hard-core bosons can be mapped to the spin-1/2-XY antiferromagnet 
\cite{Eckardt2010} with possible spin-liquid ground states in the 
spatially anisotropic triangular lattice \cite{Schmied2008etal} and
still unexplored behavior in the kagom\'e geometry. The ability to 
tune continuously between zero and maximum frustration described here can, 
thus, be a powerful tool for the adiabatic preparation of frustrated 
quantum phases. 

\begin{figure}
    \centering
\includegraphics[width=\linewidth]{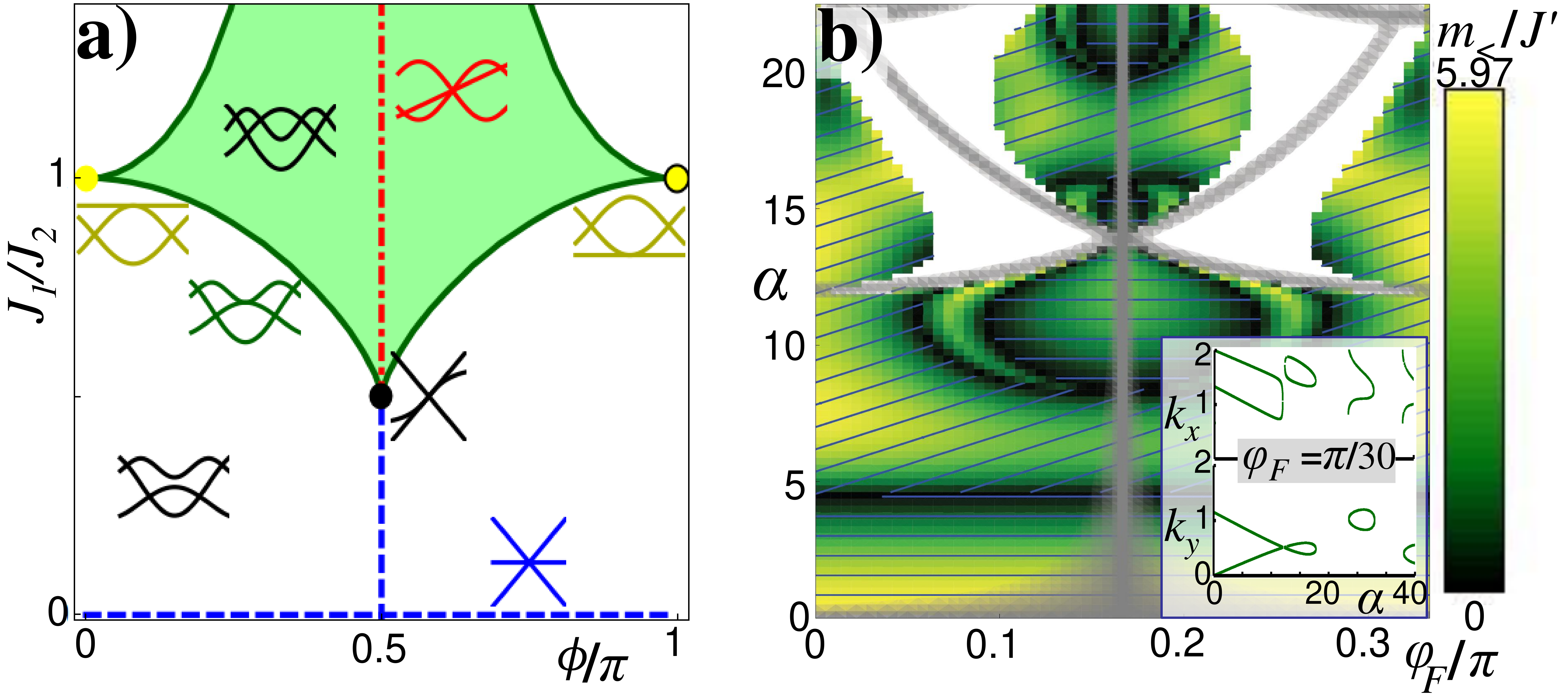}
    \caption{\label{fig:top} (color online)
{(a)} The topology of band touching for the kagom\'e lattice can be
controlled by anisotropy $J_1/J_2$ and plaquette flux $\phi$.  
The way and how often the three bands touch is depicted by the iconographic 
symbols. 
{(b)} Phase diagram of the hexagonal lattice as in 
Fig.~\ref{fig:tri}c with bare/undriven (next) nearest neighbor tunneling 
matrix elements $J$ ($J'$), subjected to a symmetry-breaking force of
amplitude $\alpha$ and direction 
${\bm e}_F=\cos(\varphi_F){\bm e}_x+\sin(\varphi_F){\bm e}_y$, $F(t)$. 
white: no Dirac points are present; gray: a small nearest-neighbor
tunneling $<0.02J$ renders the physics effectively 1D. 
The colorbar encodes the masses at two Dirac points, labeled as $\left|m_<\right|\le\left|m_>\right|$. 
In the diagonally (horizontally) hatched region both masses are positive (negative). 
When the masses have opposite sign (un-hatched),
the system is a topological insulator (or a quantum spin Hall insulator for 
two spin states).
Inset: Position of Dirac points in k-space for $\varphi_F=\pi/30$, 
indicating how they move and merge with $\alpha$. }
\end{figure}

The realization of tunable staggered fluxes as shown in
Fig.~\ref{fig:tri}a and b is also interesting in its own right. In the bosonic 
case, deviations from $\pi$-flux directly map to tunable Dzyaloshinskii-Moriya 
couplings in the spin picture (see, e.g., \cite{Messio2010}). 
Furthermore, for finite flux $\phi_\Delta=\phi$, the three bands of the kagom\'e lattice feature a complex band-touching structure whose topology can be controlled by the driving. This is illustrated in Fig.~\ref{fig:top}a for a lattice with $|J^\text{eff}_{ij}|$ equal to $J_1$ ($J_2$) along the horizontal (other) bonds (see Fig.~\ref{fig:tri}b).

\paragraph*{Topological and quantum spin Hall insulator.} 
Such triangular plaquette fluxes can be used to engineer a topological 
insulator and a quantum spin Hall insulator. 
Consider a spin-dependent hexagonal optical lattice as sketched in Fig.~\ref{fig:tri}c, where sites of the A (B) sublattice are energetically lifted (lowered) by $\Delta E/2$ for $\up$ 
particles, and vice versa for $\dw$ particles \cite{SoltanPanahi2011}.
Let us focus on non-interacting $\up$-particles first. For substantial 
detuning $\Delta E$, we can assume that nearest-neighbor (NN) tunneling
(between A and B sites) is energetically suppressed and that next-NN (NNN) 
tunneling is relevant only between sites of the ``shallow'' A sublattice. 
Now assume that the system is driven resonantly by a time-periodic homogeneous force of frequency $\nu_{AB}\hbar\omega=\Delta E$ (with integer $\nu_{AB}$) that both 
establishes NN ACT and creates finite artificial fluxes through the triangular NNN plaquettes of the A sublattice (``+'' and ``-'' in Fig.~\ref{fig:tri}c). 
Introducing Pauli matrices $\sigma$ for the sublattice degree of freedom, 
the effective Hamiltonian in momentum representation becomes
$\Ho_\text{eff} =\sum_\bk\hat{\bm a}^\dag_{\bk} 
h(\bk)\hat{\bm a}_{\bk} $ 
where $\hat{\bm a}^\dag_{\bk}=(\aa_{A\bk},\aa_{B\bk})$ and
$h(\bk)=\Re\left(g(\bk)\right)\sigma_x -\Im\left(g(\bk)\right)\sigma_y +
g'(\bk)\frac{1}{2}\left(\mathbb{1}+ s_z \sigma_z\right)$. 
Here, $s_z=1$ and $g^{(\prime)}(\bk)\equiv
-\sum_{{\bm \delta}^{(\prime)}}J^\text{eff}_{{\bm\delta}^{(\prime)}}
\exp(\ri\bk\cdot {\bm \delta}^{(\prime)})$ with ${\bm \delta}^{(\prime)}$ 
denoting the three (six) vectors connecting an A site to its NN (NNN). 
Diagonalizing $h(\bk)$ gives the dispersion relations 
$\varepsilon_\pm(\bk)=\frac{1}{2}g'(\bk)\pm\sqrt{|g(\bk)|^2+|g'(\bk)/2|^2}$ 
for the two bands. 

Without NNN tunneling ($g'=0$), the system 
can possess a pair of band-touching points, i.e., $g(\bk_{1,2})=0$, with  
light-cone-like dispersion relation, so-called Dirac cones. A finite NNN 
$g'(\bk)$ will split the bands at these points, and the Dirac-type 
dispersion relations found near $\bk_{1,2}$ acquire finite ``masses'' 
$m_{1,2}=g'(\bk_{1,2})$. If these have opposite sign, the lowest band 
possesses a finite Chern number ($\pm1$). Then, if the lowest band is entirely filled 
with $\up$ fermions, the system is a topological insulator with quantized Hall 
conductivity and robust chiral edge modes \cite{Haldane1988} (see also 
\cite{Alba2011}). Repeating the above reasoning for $\dw$ 
particles, for which the role of A and B sites is interchanged, one obtains 
the same result, but with $s_z=-1$ and inverted Hall conductivity. Therefore, 
filling the lowest band with both $\up$ and $\dw$ particles the system becomes 
a quantum spin Hall insulator with opposite chirality for the two species 
\cite{Kane2005}.

As an example, we consider unidirectional forcing
$\bF(t)=F(t){\bm e}_F$, with  ${\bm 
e}_F=\cos(\varphi_F){\bm e}_x+\sin(\varphi_F){\bm e}_y$ and $F(t)$ as 
depicted in Fig.~\ref{fig:tri}d (with $T_1=T/2$ and $\hbar\omega=\Delta E/2$, see supplemental material \cite{supp} for an analytical expression of the resulting phases).
By varying the angle $\varphi_F$ and the forcing strength 
$\alpha=dF_0T_1/(2\pi\hbar)$ (with lattice constant $d$), 
we can access various topological quantum phase transitions, 
where at least one of the masses vanishes and changes sign (Fig.~\ref{fig:top}b). 
Thus, the lowest band can acquire a non-trivial Chern number. 
The inset shows how Dirac points can be moved and merged.  

 A way to measure the topological band structure of the system 
is given by the method of Ref.\ \cite{PriceCooper12} based on semi-classical 
wave-packet dynamics. It can be applied thanks to the adiabatic principle for 
Floquet systems \cite{BreuerHolthaus89} (see \cite{EckardtHolthaus08} for its 
application to the effective Hamiltonian).

\begin{figure}[t]\centering
\includegraphics[width = 1\linewidth]{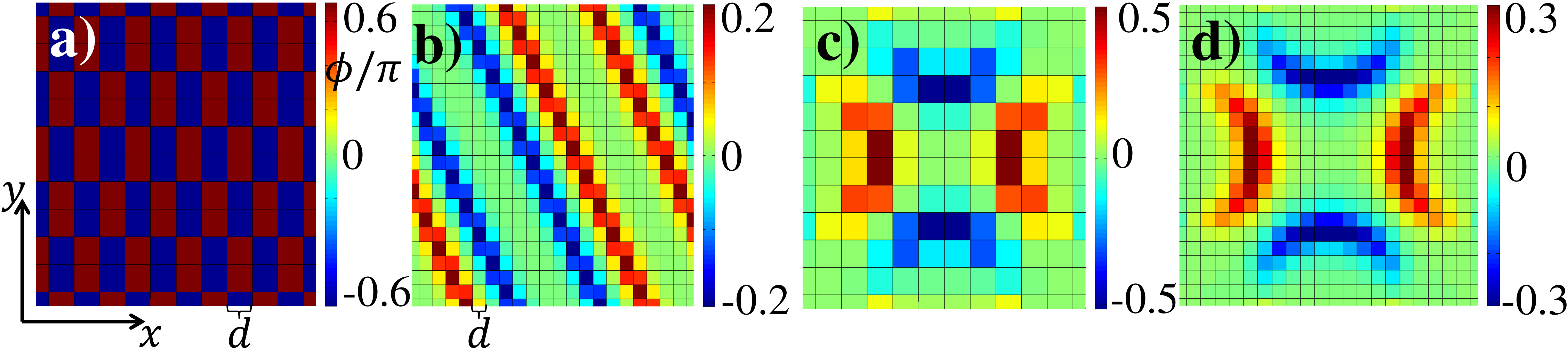}
\caption{\label{fig:super} (color online) 
Artificial magnetic fluxes $\phi$ through the plaquettes of
a square lattice (lattice constant $d$, indicated by the grid) resulting from 
superlattice modulation. Stripes or larger patches with 
strong, rectified magnetic fluxes can be achieved.
(a) and (b): Single-component superlattices with 
(a) $\bq_1=(\pi/d)({\bm e}_x + 
{\bm e}_y/2)$, $V_1=4\hbar\omega$; (b) $\bq_1=\frac{1}{10}(\pi/d)({\bm e}_x 
+ {\bm e}_y/2)$ and $V_1=20\hbar\omega$. (c): 
Two components with $\bq_{1/2}=\frac{1}{10}(\pi/d)({\bm e}_x 
\pm {\bm e}_y)$, $V_{1/2}=12\hbar\omega$. (d): Like (c), but with 
wave-lengths and amplitudes doubled. Always $\varphi_s=0$.}
\end{figure}

\paragraph*{Superlattice modulation and flux rectification.} 
In lattices with pairwise parallel bonds, such as square lattices,  
homogeneous driving $v_i^\omega(t)=-\bF(t)\cdot\br_i$ as considered in the previous paragraphs cannot create magnetic fluxes. Therefore, we propose to drive the system via an oscillating 
superlattice potential 
$	v_i(t)=f(t)V_0(\br_i) = f(t)\sum_s 
	\frac {V_s}{2}\cos(\bq_s\cdot\br-\varphi_s)$,
where $V_0(\br)$ may be incommensurate with the host lattice. 
The driving function $f(t)=f(t+T)$ breaks symmetries (r) and (s). To achieve a
vanishing mean, $\la f(t)\ra_T=0$, in an experiment one can use $\pi$-shifted
non-interfering standing waves such that $f(t)V_s 
\cos(\bq_s\cdot\br-\varphi_s)=
V_s'(t) \cos(\bq_s\cdot\br-\varphi_s)+V_s''(t) 
\cos(\bq_s\cdot\br-\varphi_s+\pi)$, with $V_s',V_s''>0$. In 
Fig.~\ref{fig:super}, we show -- on the example of a square lattice with a 
shaking function as in Fig.~\ref{fig:tri}d (with $T_1/T=0.8$) -- that, using 
different superlattice structures, various configurations of plaquette fluxes 
can be engineered \cite{footnoteLongWavelength}. Roughly, the larger the 
superlattice wavelengths the slower is the variation of the artificial flux. 
Therefore, superlattice modulation can generate not only strong magnetic 
fluxes through square plaquettes, but also large regions (stripes or patches) 
with rectified magnetic field where strong-field quantum Hall-type physics can 
be studied. Their inhomogeneity and finite extent provide a promising test 
ground for the investigation of robust edge modes.

\begin{figure}
    \includegraphics[width=\columnwidth]{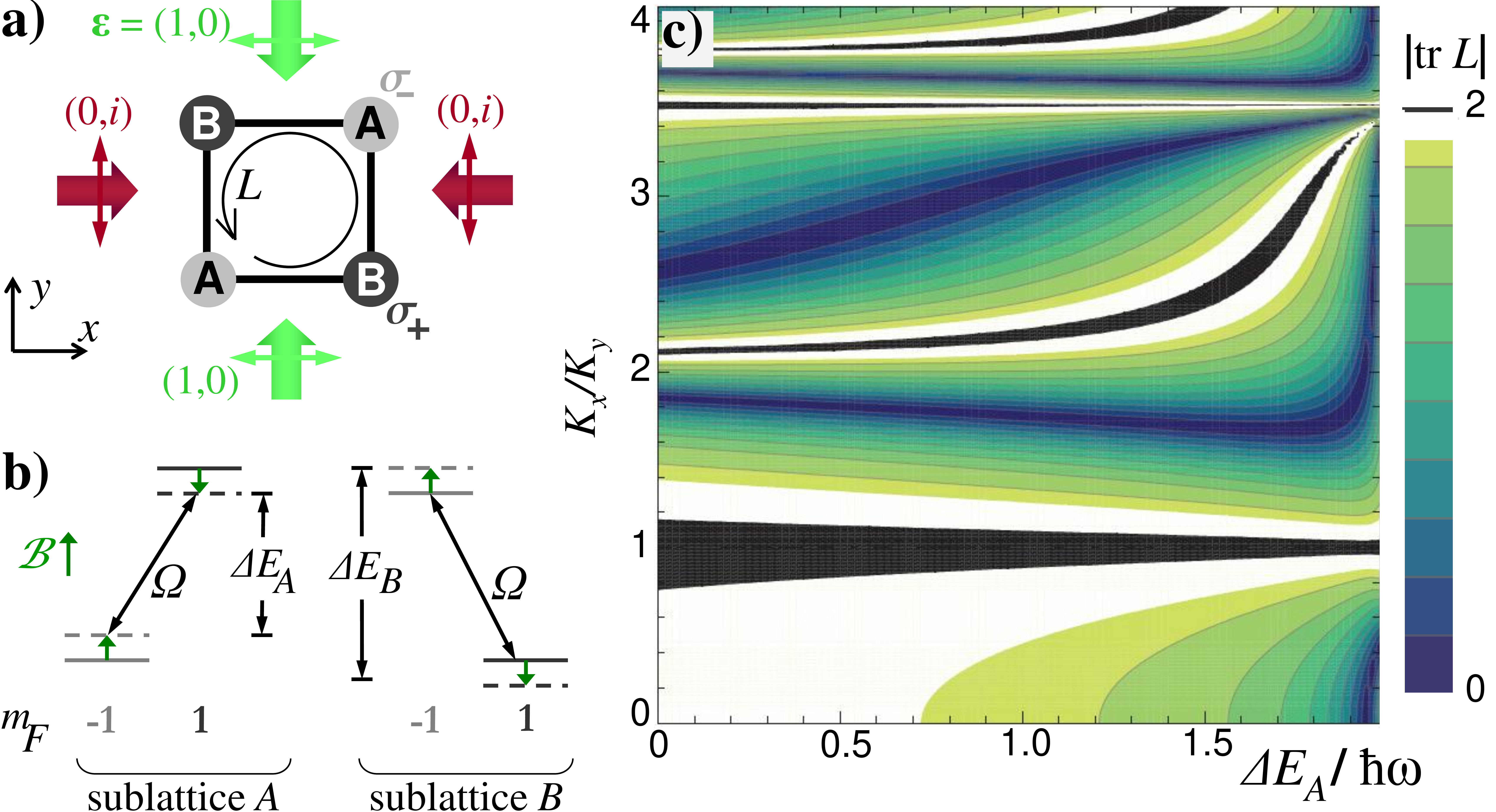}
    \caption{
(color online) Non-Abelian SU(2) gauge fields.
{(a)} 
Two standing laser waves (with a phase shift of $\pi/2$ and in-plane 
polarization as denoted in the figure) create a bipartite square lattice with 
alternating  $\sigma^+$ and $\sigma^-$ polarized sites ($A$ and 
$B$) \cite{Hemmerich1993}. $m_F=\pm1$ particles feel an energy 
difference of $\pm\Delta E$ between $A$ and $B$ sites.
{(b)} The resulting level scheme. A constant $B$-field 
realizes an additional on-site energy splitting $\Delta E'$ (green arrow)
such that $|\Delta E_{A,B}|=|\pm\Delta E+\Delta E'|$ becomes sublattice 
dependent. The coupling $\Omega$ of both spin states can be realized by 
magnetic or microwave fields.
{(c)} Trace of the Wilson loop $L$ in parameter space. Deviations from 2
imply non-Abelian physics [$K_{\hat y}=1.814$; 
outside the white (black) regions, $| \mathrm{tr} L|<1.9$ ($<1.99$)].
\label{fig:Wilson}}
\end{figure}

\paragraph*{Non-Abelian SU(2) gauge fields.}
The periodic driving also permits the creation of arbitrary \emph{non-Abelian} 
SU(2) gauge fields. Consider $\up$ and $\dw$ particles (say, $m_F=\pm1$) 
loaded into the spin-dependent square lattice depicted in Fig.~\ref{fig:Wilson}a, where the 
energy of $\up$ particles is lifted (lowered) by $\Delta E/2$ on 
A (B) sites, and vice versa for $\dw$ particles. These energy shifts are
summarized by $\Delta E\sigma_zs_z/2$, if we introduce two sets of Pauli 
matrices $s$ and $\sigma$ for spin ($\up$ or $\dw$) and sublattice ($A$ or $B$), respectively. 
Moreover, uniform microwave and magnetic fields can be employed to couple the $\up$ and $\dw$ state with a matrix element $\Omega$ and to produce an additional site-independent energy splitting $\Delta E'$, giving the site-independent term $\Delta E's_z/2+\Omega s_x$. 
The absolute value of the total $\up$--$\dw$-splitting $\Delta E_i=\Delta  E\sigma_z-\Delta E'$ is sublattice-dependent (Fig.~\ref{fig:Wilson}b).
Including the NN tunneling $J$ and a spin-independent sinusoidal drive 
$v_i^\omega(t)=-\br_i\cdot\bF_0\cos(\omega t)$ as it can be induced by simply shaking the lattice back and forth, the Hamiltonian reads 
$\Ho=-\sum_{\la ij\ra}J \hat{\bm a}^\dag_i \hat{\bm a}_j 
+ \sum_i\hat{\bm a}^\dag_i [\frac{1}{2}\Delta E_i s_z 
+ \Omega s_x + v_i^\omega(t)] \hat{\bm a}_i$, with 
$\hat{\bm a}_i^\dag=(\aa_{i\up},\aa_{i\dw})$. 
The transformation $\hat{\bm b}_i=u_i^\dag\hat{\bm a}_i$, where $u_i$ are time-independent unitary 2$\times$2-matrices, diagonalizes the Hamiltonian on site with eigenvalues $\hbar\lambda_i=\frac{1}{2} \sqrt{\Delta E_i^2+4\Omega^2}$. 
This yields 
$\Ho=-\sum_{\la ij\ra}J \hat{\bm b}^\dag_i u^\dag_i u_j \hat{\bm b}_j +
\sum_i\hat{\bm b}^\dag_i [\hbar\lambda_i s_z + v_i^\omega(t)] \hat{\bm b}_i$.
The sublattice dependence of $u_i$ through $\Delta E_i/(2\Omega)$ achieves generically $ u^\dag_i u_j \ne 1$. 
As in the derivation preceding Eq.~(\ref{eq:peierls}), the unitary transformation 
$\exp\big(-\ri\sum_i \hat{\bm b}^\dag_i [\lambda_i t s_z - K_i\sin(\omega t)] 
\hat{\bm b}_i\big)$ with $K_i=\br_i\cdot\bF_0/(\hbar\omega)$ leads to a
purely kinetic Hamiltonian $\Ho'=-\sum_{\la ij\ra}J\hat{\bm b}^\dag_i W_{ij}(t)\hat{\bm b}_j$.
Here, 
\begin{equation*}
W_{ij}(t)=
\re^{-\ri 
K_{ij}\sin(\omega t)}
\left(\begin{smallmatrix} 
                     c_{ij} \ue^{i\left(\lambda_{i}-\lambda_{j}\right)t}        & d_{ij} \ue^{-i\left(\lambda_{i}+\lambda_{j}\right)t} \\
                    -d_{ij}^\star \ue^{i\left(\lambda_{i}+\lambda_{j}\right)t} & c_{ij}^\star \ue^{-i\left(\lambda_{i}-\lambda_{j}\right)t}
             \end{smallmatrix}\right)\,,
\end{equation*}
and $c_{ij}$ and $d_{ij}$ parametrize $u^\dag_i u_j$.
For $\hbar\omega\gg J_{ij}$, we can approximate $\Ho'$ by its time average
$ 
\Ho_\text{eff}=\la\Ho'\ra_T=-\sum_{\la ij\ra}J^\text{eff}_{ij}
\hat{\bm b}^\dag_i M_{ij}\hat{\bm b}_j
$
,
with the effective tunneling matrix elements 
$J^\text{eff}_{ij}=J\sqrt{|\det(\la 
W_{ij}\ra_T)|}$, and the matrices $M_{ij}\equiv\la 
W_{ij}\ra_T/\sqrt{|\det(\la W_{ij}\ra_T)|}$.
For $J^\text{eff}_{ij}\neq0$, we require $\lambda_{i\epsilon B}\pm\lambda_{i\epsilon A}=\nu_\pm\,\omega$ with integers 
$\nu_\pm$, and for unitarity of $M_{ij}$, we require $\nu_\pm$ both either odd or even.

If the so called Wilson loop $L$, the product of the matrices $M_{ij}$ around 
a plaquette, yields not just a simple phase $\re^{\ri\phi}\mathbb{1}$ 
describing an Abelian magnetic flux $\phi$, 
the system is subjected to a genuine non-Abelian SU(2) gauge field.
This is equivalent to requiring $\left|\mathrm{tr}L\right|<2$, a \emph{sine 
qua non} for the anomalous integer quantum Hall effect \cite{Goldman2009} and 
fractional quantum Hall states with non-Abelian anyonic excitations 
\cite{Burello2010}. Without driving, $\left|\mathrm{tr}L\right|=2$, but 
including it, $\left|\mathrm{tr}L\right|<2$ can be fulfilled (see \cite{supp}).

Let us choose $\nu_+=3$ and $\nu_-=1$, achieved by $\Delta 
E_B=\sqrt{4(\hbar\omega)^2+\Delta E_A^2}$ and 
$\Omega=\sqrt{(\hbar\omega)^2-\Delta E_A/4}$. This leaves $\Delta E_A/\hbar\omega$, $K_x$, and $K_y$ as free parameters 
(where $K_{x,y}$ is the amplitude of the forcing $K_{ij}$ in positive 
$x,y$-direction). 
In Fig.~\ref{fig:Wilson}c, we plot the trace of the Wilson loop 
$\left|\mathrm{tr}L\right|$ versus $K_x/K_y$ and $E_A/\hbar\omega$ for 
$K_y=1.84$ (this value is not crucial but ensures large $y$-tunneling -- see supplemental material \cite{supp}, where also an analytical 
expression for the Wilson loop is derived). 
There are broad regions where $\left|\mathrm{tr}L\right|$ differs strongly 
from 2, proving the presence of a strong artificial non-Abelian gauge field. 
Under typical conditions, the system shows Dirac cones, be it Abelian or 
non-Abelian. Similar analytic calculations reveal that $L\equiv\mathbb{1}$ in 
a hexagonal lattice (and for even $\nu_\pm$ in the square lattice). 
This limitation can be overcome by employing position-dependent coupling via 
Raman laser mixing, $\Omega\to\Omega_{i}= \Omega
\re^{\ri\vect{q}\cdot\br_i}$ with $\vect{q}$ the laser wave-vector difference (see \cite{supp}). 
This way, the $M_{ij}$ as well as $L$ can be tuned to be a generic ($i\times$) 
SU(2) matrix both in square and hexagonal lattices. Alternatively, in a 
hexagonal lattice a non-trivial Wilson loop can be achieved with NNN 
tunneling. 

\paragraph*{Conclusion.}
The creation of artificial Abelian and non-Abelian gauge fields by means of 
time-periodic forcing opens realistic perspectives for experimental studies. 
This method offers great flexibility, because it does not involve the internal 
atomic structure. For fermions, where only different internal states interact 
with each other, this can be very advantageous for reaching the strongly 
correlated regime. 

{\bf Acknowldegments} We acknowledge support from AAII-Hubbard,
Spanish MICINN (FIS2008-00784), Catalunya-Caixa, EU Projects AQUTE and NAMEQUAM,
ERC grant QUAGATUA, Netherlands Organisation for Scientific Research (NWO), Humboldt Stiftung, German Science foundation (grants FOR 801 and SFB 925), and Hamburg Theory Prize.

\newpage

\begin{widetext}
\hspace{0.5cm}
{\bf \LARGE Supplemental Material 
}
\end{widetext}

\renewcommand{\theequation}{S\arabic{equation}}
\setcounter{equation}{0}
\renewcommand{\thefigure}{S\arabic{figure}}
\setcounter{figure}{0}

\vskip 0.5cm

\section{Analytic form of the effective tunneling and the Peierls phases}
In the examples of the main text, we use a driving potential $\bF(t)=F(t){\bm e}_F$ which is unidirectional, i.e., ${\bm e}_F=\cos(\varphi_F){\bm e}_x+\sin(\varphi_F){\bm e}_y$, and has a paused-sine-wave amplitude as depicted in Fig.~1d, 
\eq{
F(t)=\left\{
		\begin{array}{ccccccc}
                F_0\sin(2\pi t/T_1)\,, 	& \qquad &0  & \leq& t\,\mathrm{mod}\, T&<& T_1\\
		0 \,,			& \qquad &T_1& \leq &t\,\mathrm{mod}\, T&<& T
                \end{array}
		\right.
}
Carrying out the time integrations as given in the introduction of the main text, this driving creates -- for a vanishing energy difference between sites $i$ and $j$, characterized by $\nu_{ij}=0$ -- the effective tunneling 

\eq{ 
\frac{J^\text{eff}_{ij}}{J_{ij}}=
\frac{T_1}{T}\ue^{-i\gamma_{ij}}\left[-\ue^{-i\alpha_{ij}\frac{T-T_1}{T}}\mathcal{J}_{0}(\alpha_{ij}) +\ue^{i\alpha_{ij}\frac{T_1}{T}}\frac{T-T_1}{T_1}\right]
\,.
}
and for $\nu_{ij}\neq 0$ the AC-induced tunneling (ACT)
\eqa{ 
\frac{J^\text{eff}_{ij}}{J_{ij}}&=&
\frac{T_1}{T}\ue^{-i\gamma_{ij}}\left[-\ue^{-i\nu_{ij}\frac{\pi}{2}}\ue^{-i\alpha_{ij}\frac{T-T_1}{T}}\mathcal{J}_{\nu_{ij}}(\alpha_{ij}) \right.\\
&&\left.\qquad\qquad\,-\frac{i}{2\pi\nu_{ij}}\ue^{i\alpha_{ij}\frac{T_1}{T}}\left(\ue^{i 2\pi\nu_{ij}\frac{T}{T_1}}-1\right)\right] \nonumber
\,,
}
Here,  ${\cal J}_{\mu}$ is the BesselJ function of order $\mu$, $\gamma_{ij}$ correspond to the freedom to choose the local phases, and we defined the dimensionless driving amplitude $\alpha_{ij}=\vect{r}_{ij}\cdot\vect{e}_F F_0T_1/(2\pi\hbar)$. 
For the case $T_1=T/2$ (and choosing the local phases $\gamma_{ij}=0$), as considered in the section \emph{Topological and quantum spin Hall insulator}, the Peierls phases $\theta_{ij}$ are thus 
for nearest neighbors (NN) (where $\nu_{ij}=1$)
\eq{
\theta_{ij}\equiv f(\vect r_{ij})=\frac{1}{2}\left(\pi-\alpha_{ij}\right)\,,
}
and for next-NN (NNN) (where $\nu_{ij}=0$)
\eq{
\tan\theta_{ij}\equiv\tan f(\vect r_{ij})=\frac{1+\mathcal{J}_0(\alpha_{ij})}{1-\mathcal{J}_0(\alpha_{ij})}\tan\frac{\alpha_{ij}}{2}\,.
}
The flux threaded through a triangular plaquette as sketched in Fig.~1a-b 
is then $\phi=f(\vect e_{x})+f(-\vect e_{x}/2+\sqrt{3}/2 \vect e_{y})+f(-\vect e_{x}/2-\sqrt{3}/2 \vect e_{y})$ (and similarly for Fig.~1c, 
 where the plaquette is spanned by NNN tunneling). 
It can be non-zero since $f(\vect r_{ij})$ is a non-linear function of $\vect{r}_{ij}$.\\

\section{Non-Abelian SU(2) gauge fields}

\subsection{Independence of the observables on the choice of the phases of the local basis}

In the main text, it is shown how a non-Abelian gauge field can be induced via lattice shaking in a bipartite, spin-dependent lattice. Here, we comment on some technical and background aspects that have been omitted there. 
We adopt the same notation as in the main text.
First, we focus on the the example of site-independent magnetic mixing $\Omega_i=\Omega$, $\forall i$. The on-site Hamiltonian is 
\begin{equation}
H_i = E_i s_z +\Omega s_x,\label{h_0}
\end{equation}
where $E_i =E_A$, if the site $i$ belongs to sublattice $A$, while $E_i =E_B$, if it belongs to sublattice $B$. The transformation $u_i$ diagonalizing $H_i$ can be written as 
\eq{ 
\label{eq:definitionui}
u_i= \re^{\frac i2 \Lambda_i s_y} \re^{i (\varphi_i s_z + \tau_i)}\,,
}
where $\Lambda_i\equiv \atan \frac{E_i}{\Omega}$. 
The phases $\varphi_i$ and $\tau_i$ correspond to the $\gamma_i$; they attest the freedom in choosing the phases of the two states which form the local basis.
As these are arbitrary, physical observables -- for instance the Wilson Loop operator $L$ -- cannot depend on such phases. This is immediate in the absence of periodic driving. 
Indeed, in that case the Wilson loop is the identity whatever choice of $u_i$ is performed, as it corresponds to products of terms $u_i u_i^\dag=\mathbb{1}$, for any site $i$ in the loop.
In presence of periodic driving, the cancellation of the phases is slightly more involved. Following the definition of the effective hopping matrices $M_{ij}$ in the main text, 
$M_{ij}\equiv \langle W_{ij}\rangle_T/\sqrt{|\text{det}(\langle W_{ij}\rangle_T)|}$, 
we notice that that their matrix elements 
have the same phases (up to multiples of $\pi$) as the elements $u_i^\dag u_j=\left(\begin{smallmatrix} c_{ij}&d_{ij}\\-d_{ij}^\star& c_{ij}^\star\end{smallmatrix}\right)$, while their moduli are independent of such phases. 
This implies that for two different choices of the local phases at the site $k$, $\varphi_k^\prime$, $\tau_k^\prime$, and $\varphi_k$, $\tau_k$, respectively, the effective hopping matrices relate as
\begin{equation}
M_{ij}^\prime = \re^{i (\Delta \varphi_i s_z + \Delta \tau_i)} M_{ij} \re^{-i (\Delta \varphi_j s_z + \Delta \tau_j)},
\end{equation}
where $\Delta \varphi_k\equiv\varphi_k^\prime-\varphi_k$, and $\Delta \tau_k\equiv\tau_k^\prime-\tau_k$.
That is, the choice of the phases commutes with the time-average procedure. It follows that they cancel out when the hopping matrices $M_{ij}$ are multiplied in the Wilson Loop $L$ as in the time-independent case.
Hence, $L$ is independent of the choice of local phases, as it should be. 

We conclude by remarking that the same happens for more involved choices of the optical lattice, as the actual form of the local Hamiltonian does not play any role (cf.\ section \ref{raman}).           

\subsection{Analytic calculation of the Wilson Loop}

Here, we show how to derive analytically the Wilson Loop $L$ computed along the fundamental plaquette for a forced lattice described by (\ref{h_0}). Using the result explained in the previous section, we may choose $\varphi_i=\tau_i=0$, $\forall i$.
Using this in \eqref{eq:definitionui}, this implies that 
\eq{ 
u_i^\dag u_j= \re^{\frac i2 (\Lambda_j-\Lambda_i)s_y}\,.
\label{eq:uidaguj}
}
We may call $S$ the forcing-induced non-linear map that relates  $u_i^\dag u_j$  to the effective hopping matrix $M_{ij}$, $M_{ij}= S[u_i^\dag u_j]$.
Because we employed the unitary transformation $\exp\big(-\ri\sum_i \hat{\bm b}^\dag_i [\lambda_i t s_z - K_i\sin(\omega t)] 
\hat{\bm b}_i\big)$ to arrive at Eq.~(2) of the main text, this map is a function of the energy differences between A and B sites, $\lambda_{i\epsilon B}\pm\lambda_{i\epsilon A}=\nu_\pm\,\omega$. To ensure finite ACT, the $\nu_\pm$ have to be integers. 
Explicitly, as defined in the main text, $M_{ij}\equiv\la W_{ij}\ra_T/\sqrt{|\det(\la W_{ij}\ra_T)|}$, which yields, following Eq.~(2), 
\begin{equation}
\la 
W_{ij}\ra_T=
\left(\begin{smallmatrix} 
                     c_{ij} \mathcal{J}_{\nu_-}(K_{ij})        & d_{ij} \mathcal{J}_{\nu_+}(K_{ij}) \\
                     d_{ij}^\star \mathcal{J}_{\nu_+}(K_{ij}) & - c_{ij}^\star \mathcal{J}_{\nu_-}(K_{ij})
             \end{smallmatrix}\right)\,.
\end{equation}
Here, the amplitude of the forcing $K_{ij}=({\vect r}_i- {\vect r}_j)\cdot {\vect F}_0/(\hbar \omega)$ depends only on the direction $\boldsymbol{\delta}_l$ of link $ij$ 
($\boldsymbol{\delta}_{l=1,2}=\hat x,\hat y$ for the square lattice, and $\boldsymbol{\delta}_{l=1,2,3}=\hat x,-\frac 12 (\hat x - \sqrt 3 \hat y),-\frac 12 (\hat x + \sqrt 3 \hat y)$ for the hexagonal one).
To ensure unitarity $M_{ij}M_{ji}=\mathbb{1}$, we therefore have to impose the additional condition that the $\nu_\pm$ are either both even or both odd (use $\nu_-(ji)=-\nu_-(ij)$ and $K_{ji}=-K_{ij}$).

First, we characterize the action of $S$ on Eq.~\eqref{eq:uidaguj} for {\it even} forcing, i.e., when both $\nu_+$ and $\nu_-$ are even numbers, $\nu_+=2n$, $\nu_-=2n'$.  
By the observation that $S$ maps by construction {\sl i)} unitary matrices to unitary matrices, 
and {\sl ii)} real matrices to real matrices, it follows that $S[\re^{i\varphi s_y}]=\re^{i\varphi^\prime s_y}$, where $\varphi^\prime=  \atan[\frac{{\cal J}_{2n'}({K}_{ij}) }{{\cal J}_{2n}({K}_{ij})}\tan \varphi]$. 
The main consequence of this form of $S[\re^{i\varphi s_y}]$ is that hopping matrices $M_{ij}=M_l$ commute also for different links $l$, and this implies that the Wilson loop for the square, $L=M_1 M_2 M_1^\dag M_2^\dag$, and for the hexagonal lattice
$L=M_1 M_3^\dag M_2 M_1^\dag M_3 M_2^\dag$, are both equal to the identity (by convention, we travel the loop on fundamental cells anticlockwise, starting from the bottom-left corner).

The situation is more interesting for {\it odd} forcing, i.e., when both $\nu_+$ and $\nu_-$ are odd numbers, $\nu_+=2n+1$, $\nu_-=2n'+1$. The result of the time-averaging is in this case
   $S[\re^{i\varphi s_y}]=s_z\re^{i\varphi^\prime s_y}$, where $\varphi^\prime=  \atan[\frac{{\cal J}_{2n'+1}({K}_{ij}) }{{\cal J}_{2n+1}({K}_{ij})}\tan \varphi]$. 
   Due to the presence of the $s_z$, for the square lattice, starting by convention from sublattice  $A$, we get
\begin{align*}
 L&=M_1 M_2 M_1^\dag M_2^\dag= \cr
  &=s_z \re^{i\varphi'_x s_y} s_z \re^{i\varphi'_y s_y} \re^{-i\varphi'_x s_y} s_z \re^{-i\varphi'_y s_y} s_z=\re^{i(\varphi'_y-\varphi'_x) s_y},
\end{align*}
where 
$\varphi'_\mu = \atan\left[\frac{{\cal J}_{2n'+1}(K_\mu)}{{\cal J}_{2n+1}(K_\mu)} \tan[\frac 12(\atan \frac {E^B}{\Omega} -\atan \frac {E^A}{\Omega})]\right]$, $\mu=x,y$. 
As shown numerically in the main text, the difference $\varphi'_y-\varphi'_x$ can take any value between zero and $2\pi$, allowing for a non-trivial Wilson Loop, $\mathrm{tr}|L|= 2 \left|\cos(\varphi'_y-\varphi'_x)\right|\neq 2$.
In the example, we have chosen $K_y=1.84$, because this lies close to the first maximum of the BesselJ function ${\cal J}_{1}(K_y)$, ensuring large tunneling $J^\text{eff}_{ij}=J\sqrt{|\det(\la W_{ij}\ra_T)|}$ in $y$-direction. 

On the other hand, for the hexagonal lattice we get always a trivial result as the $s_z$'s cancel out and $L=\mathbb{1}$.

\subsection{Raman-induced mixing}\label{raman}

 \begin{figure}
\vspace*{-9cm}
\includegraphics[width=13cm]{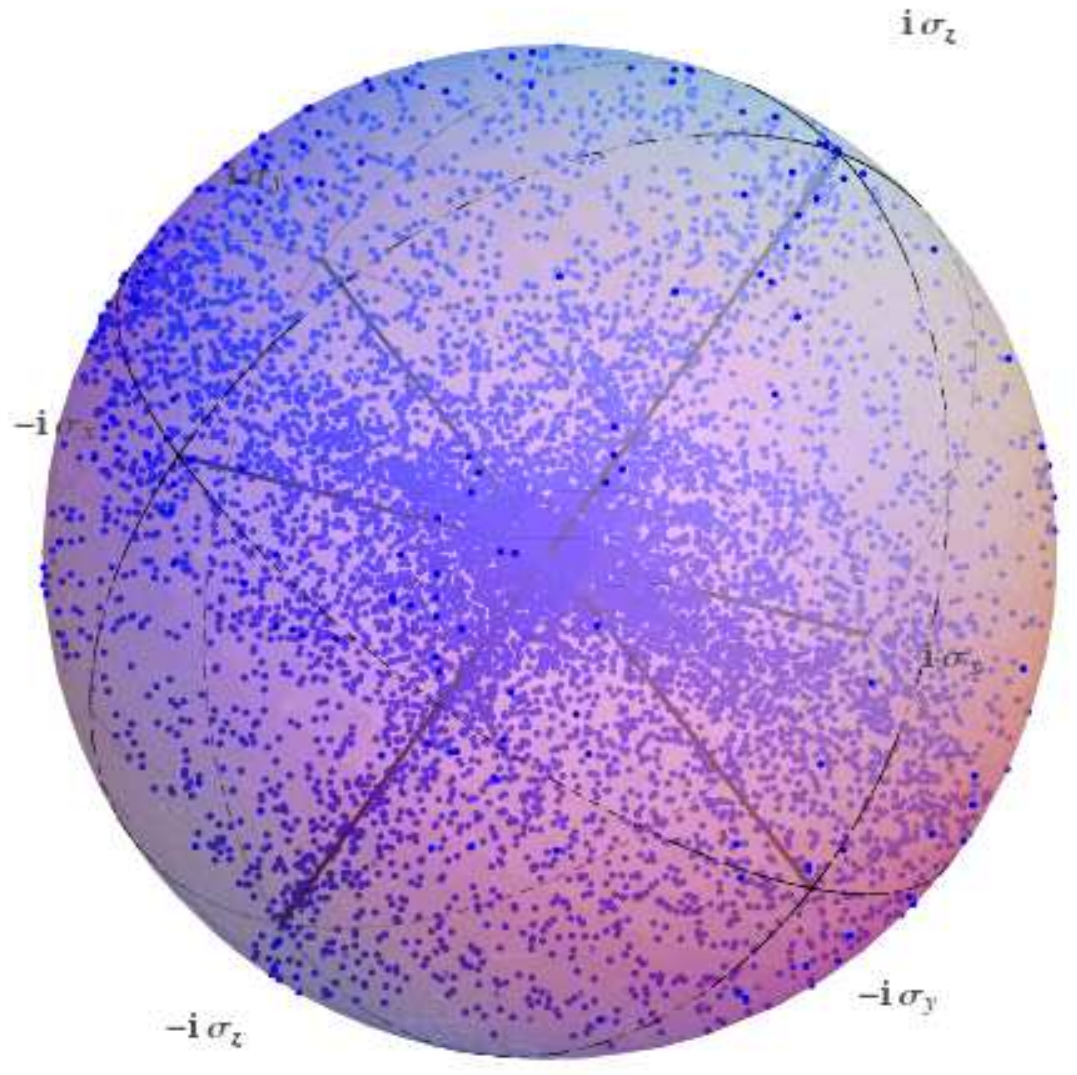}
 \caption{Bloch sphere representation of the Wilson loop $L$ computed for odd forcing in a hexagonal lattice with $|{\vect q}|=1$ (in lattice step units). 
 $E_B$ is taken to be $E_B=-\sqrt{4 E_A^2+3\Omega }$ such that the eigenvalues enjoy the relation $\lambda_B=2\lambda_A=\sqrt{E_A^2+\Omega }=2\omega$, 
 which implies $\lambda_B+\lambda_A=3\omega$, and $\lambda_B-\lambda_A=\omega$. 
 Here, $\omega$ is the frequency of the periodic forcing. 
 The parameters ${\vect q}=(\cos \theta_q,\sin \theta_q)$, the intensity of the shaking ${\vect K}=|{\vect K}| (\cos \theta_K,\sin \theta_K)$, and $E_A$, 
 are chosen randomly in the ranges, $\theta_q\in[0,2\pi[$, $|{\vect K}|\in ]0,5[$, $\theta_K\in[0,2\pi[$, and $\frac{E_A}{\Omega} \in ]5,15[$, respectively. 
The resulting $L$'s cover the entire Bloch sphere.
\label{fig:bshexq1} }
\end{figure}

Considering a site-independent mixing of $\uparrow$- and $\downarrow$-particles as in (\ref{h_0}), the form of $M_{ij}$ is limited to $\re^{i\varphi^\prime s_y}$ or $s_z\re^{i\varphi^\prime s_y}$. 
This can be circumvented by considering site-dependent Raman mixing. In this case, the local Hamiltonian takes the form 
\begin{equation}
H_i= E_i s_z + \Omega\left[ \cos({\vect q}\cdot {\vect r}_i) s_x -  \sin({\vect q}\cdot {\vect r}_i) s_y\right],\label{h1}
\end{equation}    
 where $\vect q$ is the difference of the wave vector of the two Raman lasers, $E_i=E_A$, $\forall i\in A$, and $E_i=E_B$, $\forall i\in B$.
 
 The main difference in this case is that, while still depending only on the link direction ${\vect r}_i-{\vect r}_j$, the $M_{ij}$ for different link directions are generally not commuting. Indeed, in this case the local transformation $u_i$ may be chosen as
 $u_i=\re^{i\frac{{\vect q} \cdot {\vect r}_i}2 s_z}\re^{\frac i2 \atan \frac{E_i}{\Omega}s_y}$. Hence, considering a site $j$ and all its neighbors $i$,  
 the products $u_i^\dag u_j$ for different link directions are not commuting, and the map $S$ acts highly non-trivial on them.
 Numerical studies with random selected parameters (Fig.~\ref{fig:bshexq1}) show that the corresponding Wilson Loop $L$ is in general non-trivial and dense in the Bloch sphere, i.e., we can simulate any designed $L\in $ SU(2), 
 not just matrices of the form $\re^{i\theta s_y}$.


\begin{thebibliography}{10}
\bibitem{footnoteLongWavelength}
For long superlattice wave lengths $|\bq_s d|\ll\pi$ (applies to  Fig.~\ref{fig:super}b,c,d), the flux through a plaquette at $\br$ is given by $\phi(\br) \simeq \frac{1}{2}w_{xy}(\br)\left[\theta'\left(w_y(\br)\right) -\theta'\left(w_x(\br)\right) \right]$, where a prime indicates the derivative, and where $w_{\mu\nu}(\br)=d^2\partial_\mu\partial_\nu V(\br)/(\hbar\omega)$ and $w_{\mu}=d\partial_{\mu} V(\br)/(\hbar\omega)$.
\end{thebibliography}
\end{document}